\documentclass[epj,nopacs]{svjour}

\usepackage{graphicx}
\usepackage{amsmath}

\newcommand*{\hc}[1]{{#1}^{\dag}}
\newcommand*{\angled}[1]{\left\langle #1 \right\rangle}
\newcommand*{\zetamax}{{\zeta}_{\mathrm{max}}}
\newcommand*{\ket}[1]{{\left| #1 \right\rangle}}
\newcommand{\braket}[2]{{\left\langle #1 \middle| #2 \right\rangle}}
\newcommand*{\abs}[1]{\left| #1 \right|}
\newcommand{\ee}{\mathrm{e}}
\newcommand{\ii}{\mathrm{i}}
\newcommand{\dd}{\mathrm{d}}
\newcommand{\quantity}[3]{#1\times 10^{#2}\;#3}
\newcommand{\per}[1]{#1^{-1}}
\newcommand{\second}{\text{s}}
\newcommand{\set}[1]{\{#1\}}
\newcommand{\av}[1]{{\left\langle #1 \right\rangle}}

\begin{document}

\title{The Dynamics of a Polariton Dimer in a Disordered Coupled Array of Cavities}
\author{Abuenameh Aiyejina \and Roger Andrews}
%
%
\institute{The Department of Physics, The University of the West Indies, St. Augustine, Trinidad and Tobago}
\date{}
%
\abstract{
We investigate the effect of disorder in the laser intensity on the dynamics of dark-state polaritons in an array of 20 cavities, each containing an ensemble of four-level atoms that is described by a Bose-Hubbard Hamiltonian. We examine the evolution of the polariton number in the cavities starting from a state with either one or two polaritons in one of the cavities. For the case of a single polariton without disorder in the laser intensity, we calculate the wavefunction of the polariton and find that it disperses away from the initial cavity with time. The addition of disorder results in minimal suppression of the dispersal of the wavefunction. In the case of two polaritons with an on-site repulsion to hopping strength ratio of 20, we find that the polaritons form a repulsively bound state or dimer. Without disorder the dimer wavefunction disperses similarly to the single polariton wavefunction but over a longer time period. The addition of sufficiently strong disorder results in localization of the polariton dimer. The localization length is found to be described by a power law with exponent -1.31. We also find that we can localise the dimer at any given time by switching on the disorder.
}
\maketitle

\section{Introduction}
\label{sec:Introduction}

Bound states where the composite object has a lower energy than its separated constituents due to an attractive interaction are ubiquitous. On the other hand, bound states that have a greater energy due to a repulsive interaction are rarer. Such bound states have been experimentally demonstrated by Winkler et al.~\cite{Winkler:2006tg} using repulsively interacting atoms in an optical lattice. These repulsively bound states have been theoretically studied using the Bose-Hubbard model~\cite{Valiente:2008qe,Javanainen:2010zh,Petrosyan:2007rr,Wang:2008rr}. Valiente and Petrosyan~\cite{Valiente:2008qe} were able to use the Bose-Hubbard model with a repulsive on-site interaction to derive explicit expressions for the energies and wavefunctions of the possible two-particle states and found that there were both scattering states and bound states. Javanainen, Odong and Sanders~\cite{Javanainen:2010zh} also derived expressions for the energies and wavefunctions of both scattering and bound states. They also proposed three ways to detect the dimer: by measuring the momentum distribution of the atoms, by finding the size of the molecule with measurements of atom number correlations at two lattice sites, and by dissociating a bound state of the lattice dimer with a modulation of the lattice depth. The momentum distribution was considered in the context of a possibly large number of dimers in an optical lattice when the lattice is removed and the atomic cloud expands ballistically. This method was chosen to closely resemble the actual laboratory experiments~\cite{Winkler:2006tg}. The calculation of the dissociation rate of the dimer was also linked to the laboratory experiment and the results obtained by Javanainen, Odong and Sanders were found to qualitatively explain the line shapes found by Winkler et al. In contrast to the previous two methods which dealt with large numbers of dimers, the method of using the size of the bound state applied only to the case of exactly two atoms in the lattice. The experiment must be carried out multiple times and the detection statistics compiled.

Petrosyan et al.~\cite{Petrosyan:2007rr} were able to derive an effective Hamiltonian for a lattice loaded only with dimers and were able to demonstrate theoretically that the dimer-dimer interaction has strong on-site repulsion and nearest-neighbour attraction. Wang et al.~\cite{Wang:2008rr} showed theoretically that initially prepared atom pairs are dynamically stable when the on-site repulsion is much stronger than the hopping strength. Santos and Dykman~\cite{Santos:2012et} considered a Bose-Hubbard model with a single defect with excess energy and found that even when energy conservation would allow a bound pair to dissociate into a particle localised at the defect and a particle propagating through the lattice, the bound pairs are stable against dissociation due to a destructive quantum interference.

Valiente and Petrosyan~\cite{Valiente:2008mb} theoretically examined the effect of adding a weak parabolic potential to the Bose-Hubbard model and found different behaviours between attractively and repulsively bound dimers that they attribute to differences in the ground states corresponding to attractive and repulsive on-site interactions.
Bound states have also been theoretically investigated in the extended Bose-Hubbard model~\cite{Valiente:2009pb,Wang:2010kl}. Valiente and Petrosyan~\cite{Valiente:2009pb} investigated the effect of adding nearest-neighbour interactions to the Bose-Hubbard model and found that depending on the strength of the on-site and nearest-neighbour interactions, the system possesses either one or two families of bound states, as well as scattering resonances when the bound and scattering states are degenerate. Wang et al.~\cite{Wang:2010kl} examined the effect of adding two-body nearest-neighbour interactions, including atom-pair tunnelling, to the Bose-Hubbard model. They also found the existence of either one or two bound-state solutions. For sufficiently large on-site repulsion, they found that the addition of a pair-hopping term causes a drastic change in the energy spectrum of the bound pair as compared to the standard Bose-Hubbard model. They also found that the pair-hopping term leads to metastable bound pairs with finite lifetimes.

Sanders et al~\cite{Sanders:2011kc} used a two-channel Bose-Hubbard model that explicitly includes molecules through an atom-molecule conversion operator in the Hamiltonian. They found that this leads to two bound states that represent attractively and repulsively bound dimers occurring simultaneously.

Deuchert et al.~\cite{Deuchert:2012oz} theoretically investigated the dynamics of two bosons in an infinite one-dimensional optical lattice for initial states with different separations between the two bosons. They found that for an initial state with both bosons on the same site, the bound state becomes more stable as the on-site interaction is increased. The stability of the bound state was measured using the evolution of the pair probability, which is the probability of finding both bosons on the same site. Meanwhile, when the bosons are initially separated, there is an optimal interaction strength that gives maximum pair formation.

Zhang et al.~\cite{Zhang:2013ec} examined the effect of adding an attractive defect potential to the two-particle Bose-Hubbard model. They found that for a repulsive on-site interaction, there is a critical value of the on-site interaction strength below which the ground state has the two particles localised at the defect. There is also a second, higher critical value above which the ground state has the two particles bound at the defect by the on-site repulsion. Between the two critical values, the ground state has one particle localised at the defect while the other moves through the lattice. They also investigated the effect of an attractive on-site interaction and determined that it allows the possibility of a bound state whose energy lies within the continuum band~\cite{Zhang:2012pi}.

One method of engineering quantum many-body phenomena involves using coupled optical cavities. In the case of an array of coupled cavities, each interacting with a two-level system, the system can be described by the Jaynes-Cummings-Hubbard (JCH) model~\cite{Angelakis:2007a}. This model can be implemented using quantum dots that are embedded in photonic crystals~\cite{Badolato:2005a,Hennessy:2007a,Notomi:2008a,Busch:2007a}. Angelakis et al.\cite{Angelakis:2007a} showed that a photon blockade in coupled optical cavities interacting with two-level systems can lead to a Mott-insulator phase of polaritons and its transition to a superfluid of photons. They also found that it can lead to the realization of $XY$ spin models. Reviews of theoretical and experimental advances in the understanding of coupled optical cavities have been done by Hartmann et al.~\cite{Hartmann:2008sw}, Chang et al.~\cite{Chang:2014a} and Hartmann~\cite{Hartmann:2016a}. The phase diagram of the JCH model has been obtained using mean-field theory by Greentree et al.~\cite{Greentree:2006pr}. Rossini and Fazio~\cite{Rossini:2007ud} used numerical techniques to go beyond mean-field and obtain the phase diagram for the JCH model. They also obtained the phase diagram for a system of coupled cavities, each containing a number of four-level atoms that are driven by an external laser. They found that for the phase diagram, the Bose-Hubbard model is a good approximation for the system of four-level atoms as long as the number of atoms in each cavity is sufficiently large. The system of four-level atoms has been investigated by Hartmann et al.~\cite{Hartmann:2006sj} for the case of a large number of atoms in each cavity. When the parameters of the system are suitably chosen, this system can be described in terms of a Bose-Hubbard model for dark-state polaritons. Hartmann et al. showed theoretically that this polariton system exhibits a Mott-insulator to superfluid transition. Additionally, they have shown that a two-component Bose-Hubbard model can be implemented by tuning the parameters of the system~\cite{Hartmann:2008tw}.

With respect to the polariton system examined by Hartmann et al., the effect of non-uniform laser intensities on the phase diagram and dynamics of the system was investigated by Aiyejina and Andrews~\cite{Aiyejina:2016dp} using a Gutzwiller approach. For a one-dimensional system of 25 cavities, disorder in the laser intensity was found to lead to the appearance of a Bose-glass phase in the phase diagram. The evolution of the system from its ground state while subject to a ramp in the intensity of the laser was also considered. The addition of disorder in the laser intensity was found to affect the behaviour of the defect density, superfluid order parameter and excess energy pumped into the system. In another paper, the effect of disorder in the laser intensity on the dynamics of three polaritons in a three-cavity system was considered~\cite{Aiyejina:2017fe}. The addition of disorder did not affect the general behaviour of correlation functions between the cavities, but it did suppress oscillations in those functions.

In this paper, we will examine the effect of disorder in the laser intensity on the dynamics of both a single polariton and a polariton dimer in a one-dimensional array of 20 cavities. In Section~\ref{sec:Theory} we introduce the Bose-Hubbard Hamiltonian describing the polariton system and show its extension to a time-dependent, disordered Rabi frequency. In Section~\ref{sec:Method} we outline the methods used to perform the simulations and in Section~\ref{sec:Results} we report the results obtained for the time evolution of the polariton number in the cavities for initial states of either one or two polaritons in a single cavity. The results were obtained for a uniform Rabi frequency and for uniformly distributed disorder in the Rabi frequency.

\section{Theory}
\label{sec:Theory}

The Bose-Hubbard Hamiltonian for bosons on a lattice is given by
\begin{equation}
H = \frac{1}{2}U\sum_{i}{n_i\left(n_i - 1\right)} - J\sum_{\angled{ij}}{\left(\hc{a}_i a_j + \hc{a}_j a_i\right)},\label{eq:BH}
\end{equation}
where $\hc{a}_i$ ($a_i$) is the creation (annihilation) operator for bosons at lattice site $i$, $n_i = \hc{a}_i a_i$ is the number operator for bosons at lattice site $i$ and $\angled{ij}$ indicates a sum over pairs of adjacent sites $i$ and $j$. Here, $U$ is the strength of the on-site repulsion between bosons at a given site and $J$ is the strength of the hopping of bosons between adjacent sites.

\begin{figure}
\begin{center}
\includegraphics[width=\linewidth]{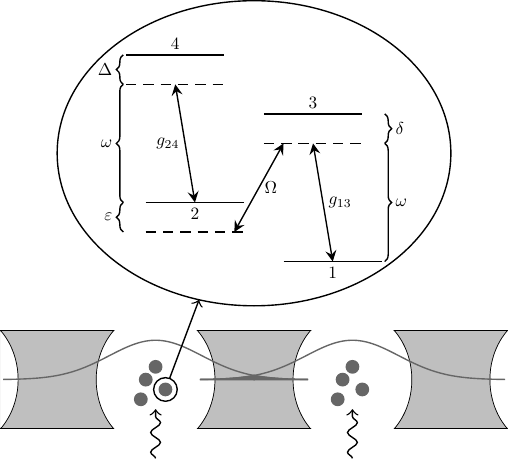}
\caption{(Above) The level structure of the atoms in the cavities. (Below) The optical cavities have electromagnetic modes that overlap for adjacent cavities. The atoms in each cavity are driven by external lasers.}
\label{fig:schematic}
\end{center}
\end{figure}

Hartmann et al.~\cite{Hartmann:2006sj} described a polariton system consisting of an array of optical cavities, each containing an ensemble of $N$ four-level atoms with the level structure shown in Figure~\ref{fig:schematic}. The cavities are sufficiently close that there is an overlap of the evanescent electromagnetic fields of adjacent cavities as also shown in Figure~\ref{fig:schematic}. Each cavity has a resonance frequency $\omega$, and the overlap integral for the electromagnetic modes of adjacent cavities is given by $\alpha$. The strengths of the couplings between the electromagnetic mode in a cavity and the transitions between atomic levels 1 and 3 and levels 2 and 4 are given by the dipole coupling parameters $g_{13}$ and $g_{24}$ respectively. An external laser drives the transition between levels 2 and 3 with Rabi frequency $\Omega$. The detunings from levels 2, 3 and 4 are given by $\varepsilon$, $\delta$ and $\Delta$ respectively. When the couplings, Rabi frequency and detunings satisfy certain conditions, and cavity and atomic decay are neglected, the system can be described by a Bose-Hubbard model for dark-state polaritons that is given by \eqref{eq:BH}. In this case, the on-site repulsion and hopping strength are given by
\begin{equation}
U = -\frac{2g_{24}^2}{\Delta}\frac{g^2\Omega^2}{\left(g^2 + \Omega^2\right)^2}
\quad \mathrm{and} \quad
J = \frac{2\omega\alpha\Omega^2}{g^2 + \Omega^2}
\end{equation}
where $g = \sqrt{N}g_{13}$. In the case of a time-dependent, non-uniform Rabi frequency, $\Omega_i(t)$, the Hamiltonian becomes
\begin{equation}
H(t) = \frac{1}{2}\sum_{i}{U_i(t)n_i\left(n_i - 1\right)} - \sum_{\angled{ij}}{J_{ij}(t)\left(\hc{a}_i a_j + \hc{a}_j a_i\right)},\label{eq:BHt}
\end{equation}
where the cavity-dependent and time-dependent on-site repulsion and hopping strengths are given by
\begin{equation}
U_i(t) = -\frac{2g_{24}^2}{\Delta}\frac{g^2\Omega_i(t)^2}{\left(g^2 + \Omega_i(t)^2\right)^2}
\end{equation}
and
\begin{equation}
J_{ij}(t) = \frac{2\omega\alpha\Omega_i(t)\Omega_j(t)}{\sqrt{g^2 + \Omega_i(t)^2}\sqrt{g^2 + \Omega_j(t)^2}}
\end{equation}
respectively. In this paper, we use uniformly distributed disorder in the Rabi frequency and we consider a case where the disorder strength can change at different times. For this uniform disorder, the Rabi frequency at cavity $i$ is given by $\Omega_i(t) = \Omega(t)(1 + \zetamax(t)x_i)$, where $\Omega(t)$ is the time-dependent mean Rabi frequency, $\zetamax(t)$ is the time-dependent disorder strength, and $x_i$ is an uncorrelated random variable uniformly distributed in the interval $\left[-1, 1\right]$.

\section{Method}
\label{sec:Method}

In this paper, we considered the dynamics of both a single polariton and two polaritons in a one-dimensional lattice of 20 cavities with open boundary conditions. For the single polariton case, we worked in the subspace of states with a total of 1 polariton. The basis of this subspace is given by $\set{\ket{n_1,\ldots,n_{20}} \mid \sum_{i}n_i = 1}$. We used an initial state of $\ket{\psi(0)} = \ket{0,\ldots,0,1,0,\ldots,0}$, where there is 1 polariton in cavity 10. For the two polariton case, we worked in the subspace of states with a total of 2 polaritons. The basis of this subspace is given by $\set{\ket{n_1,\ldots,n_{20}} \mid \sum_{i}n_i = 2}$. We used an initial state of $\ket{\psi(0)} = \ket{0,\ldots,0,2,0,\ldots,0}$, where there are 2 polaritons in cavity 10. Using the relevant basis, the Hamiltonian matrix was constructed and the Schr\"odinger equation $\ii\frac{\dd}{\dd t}\ket{\psi(t)} = H(t)\ket{\psi(t)}$ was expanded as a system of ordinary differential equations for the time-dependent elements of the vector $\ket{\psi(t)}$. This system was then solved using a Runge-Kutta method. 

\section{Results}
\label{sec:Results}

For the simulations in this paper, we used the parameters $g_{13} = g_{24} = \quantity{2.5}{9}{\per\second}$, $\varepsilon = 0$, $\delta = \quantity{1.0}{12}{\per\second}$, $\Delta = \quantity{-2.0}{10}{\per\second}$, $N = 1000$, and $2\omega\alpha = -\quantity{1.1}{7}{\per\second}$. These parameters correspond to numerical calculations for caesium atoms in toroidal microcavities~\cite{Spillane:2005xz}. The Rabi frequency was set to $\Omega = \quantity{1}{11}{\per\second}$. This gives a hopping strength of $J \approx \quantity{6.8}{6}{\per\second}$ and an on-site repulsion of $U \approx \quantity{1.5}{8}{\per\second}$. This corresponds to a value of $U/J \approx 20$, which will be observed to be sufficient to bind the dimer in the two polariton case.

\begin{figure*}
	\begin{center}
		\includegraphics[width=\textwidth]{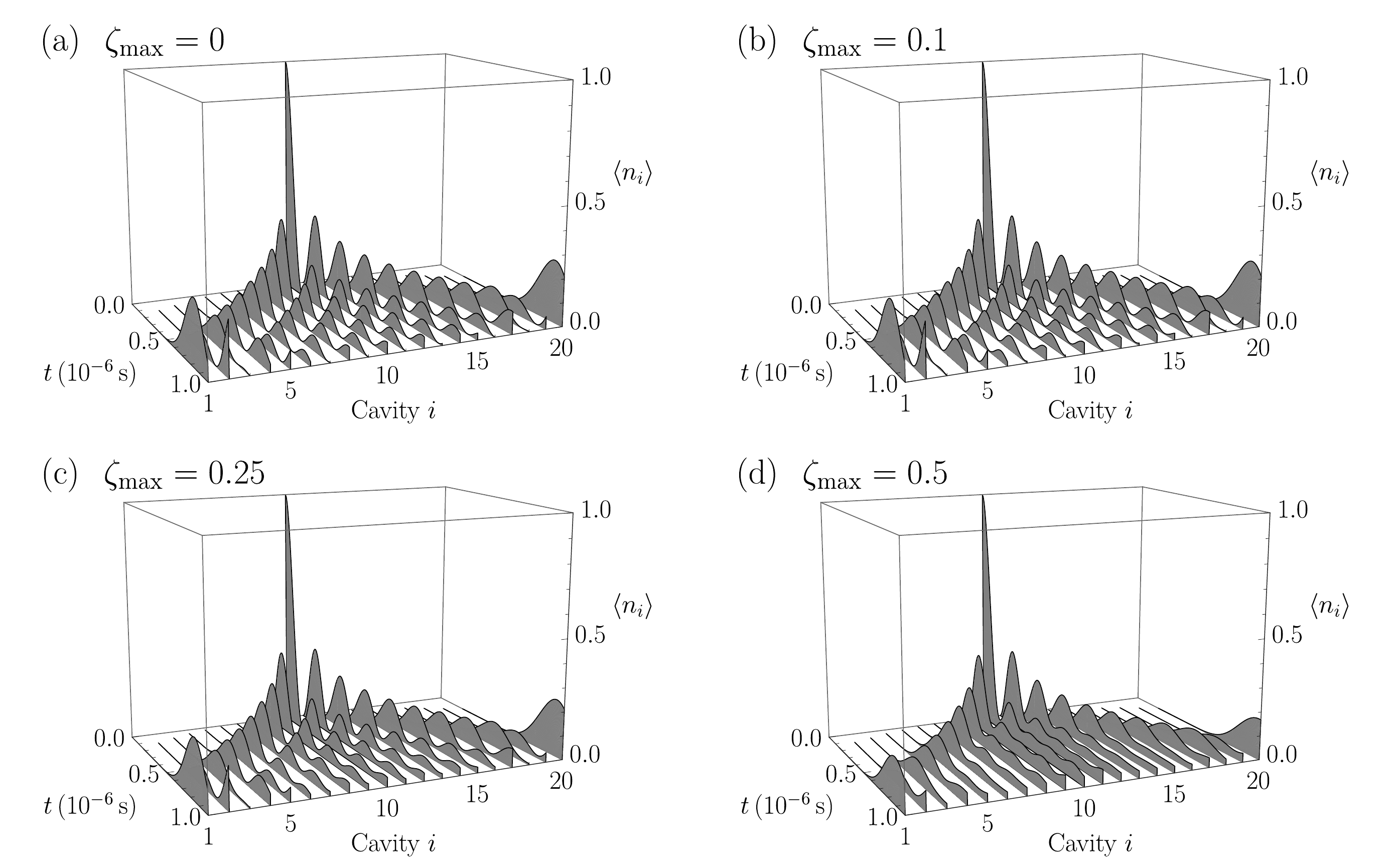}
		\caption{Dynamics of one polariton for disorder strengths (a) $\zetamax = 0$, (b) $\zetamax = 0.1$, (c) $\zetamax = 0.25$, and (d) $\zetamax = 0.5$.}
		\label{fig:mnplots}
	\end{center}
\end{figure*}

\begin{figure}
	\begin{center}
		\includegraphics[width=\linewidth]{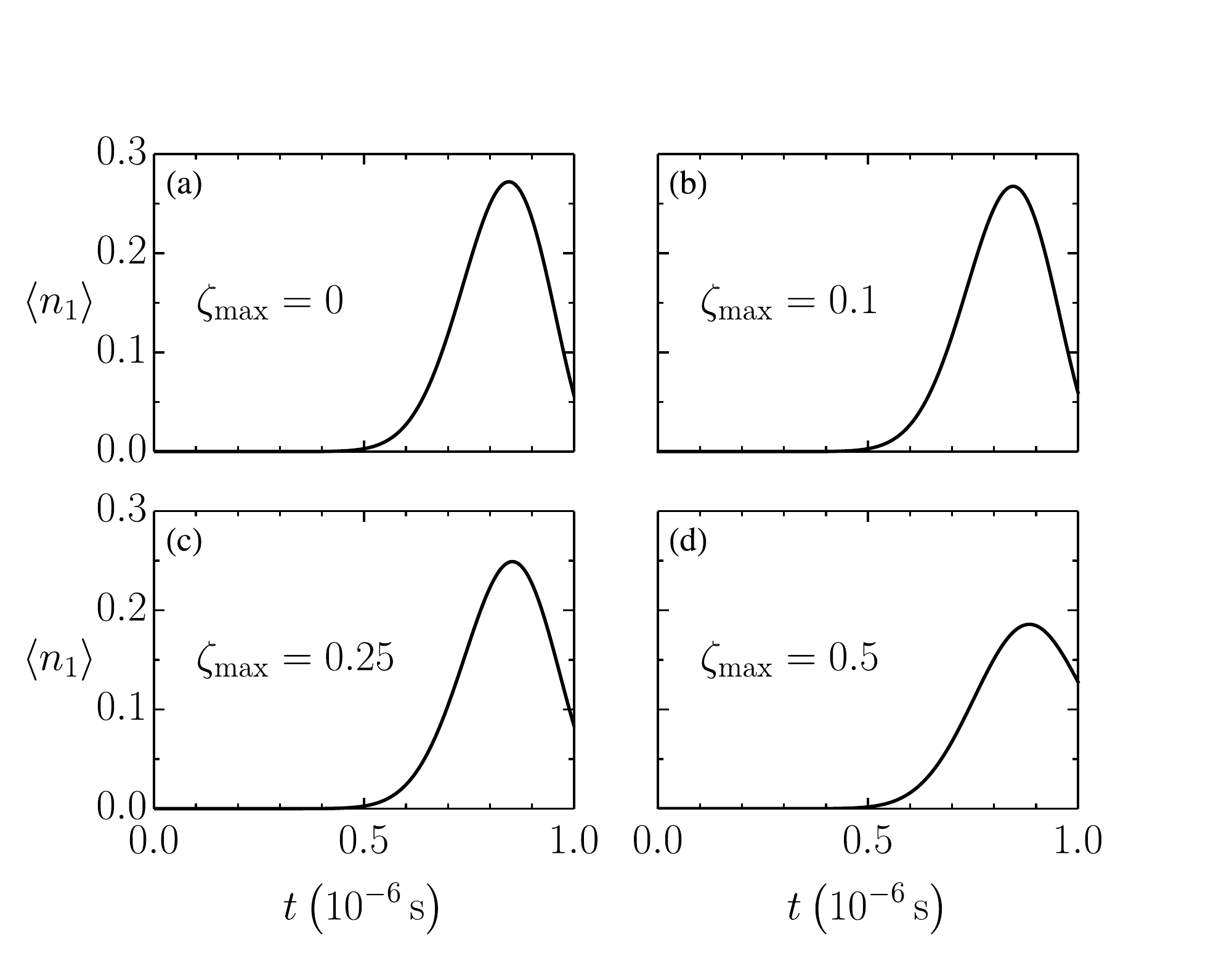}
		\caption{Evolution of the occupation of cavity 1 for disorder strengths (a) $\zetamax = 0$, (b) $\zetamax = 0.1$, (c) $\zetamax = 0.25$, and (d) $\zetamax = 0.5$.}
		\label{fig:cavnplot1}
	\end{center}
\end{figure}

\begin{figure}
	\begin{center}
		\includegraphics[width=\linewidth]{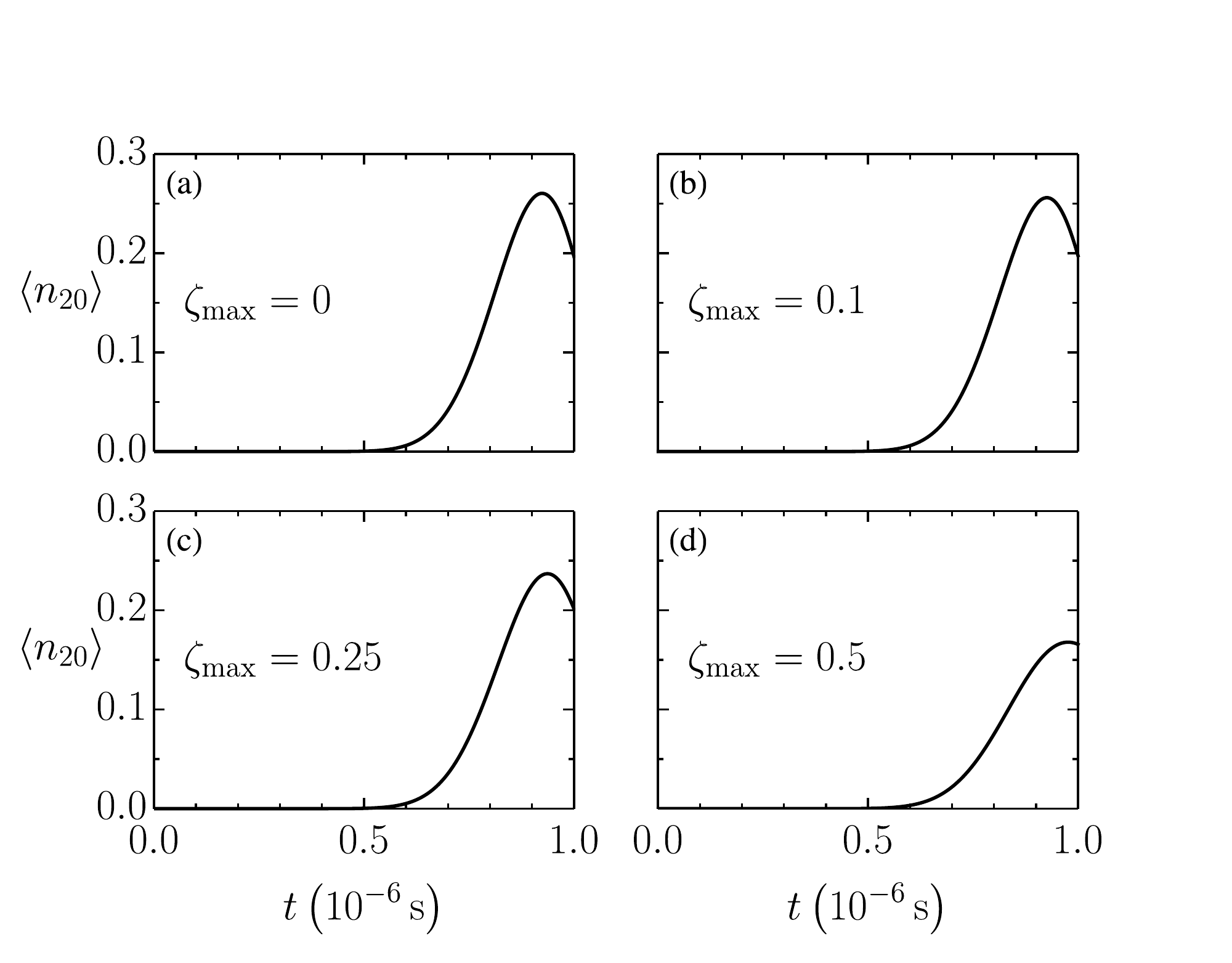}
		\caption{Evolution of the occupation of cavity 20 for disorder strengths (a) $\zetamax = 0$, (b) $\zetamax = 0.1$, (c) $\zetamax = 0.25$, and (d) $\zetamax = 0.5$.}
		\label{fig:cavnplot20}
	\end{center}
\end{figure}

\begin{figure}
	\begin{center}
		\includegraphics[width=\linewidth]{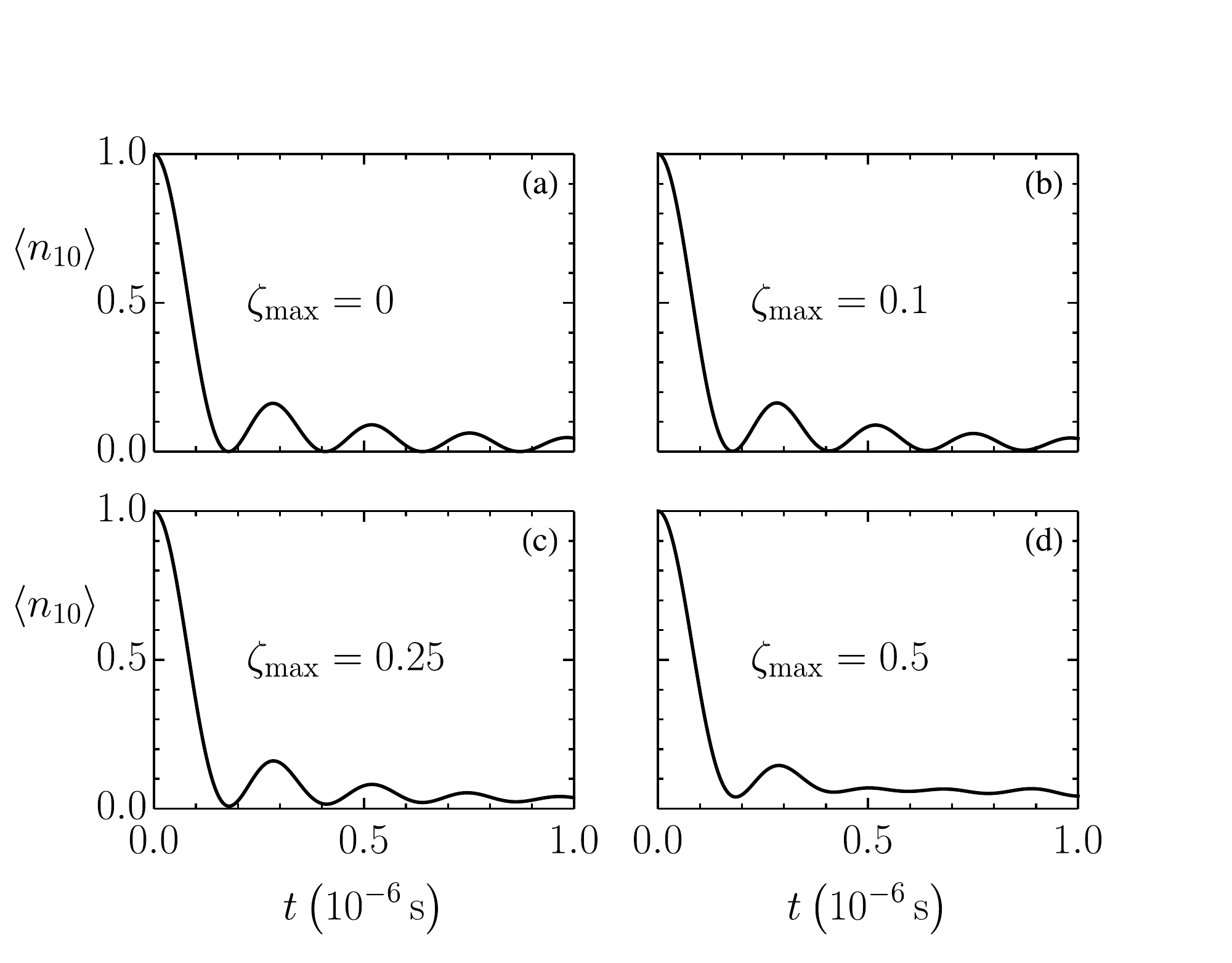}
		\caption{Evolution of the occupation of cavity 10 for disorder strengths (a) $\zetamax = 0$, (b) $\zetamax = 0.1$, (c) $\zetamax = 0.25$, and (d) $\zetamax = 0.5$.}
		\label{fig:cavnplot10}
	\end{center}
\end{figure}

\begin{figure}
	\begin{center}
		\includegraphics[width=\linewidth]{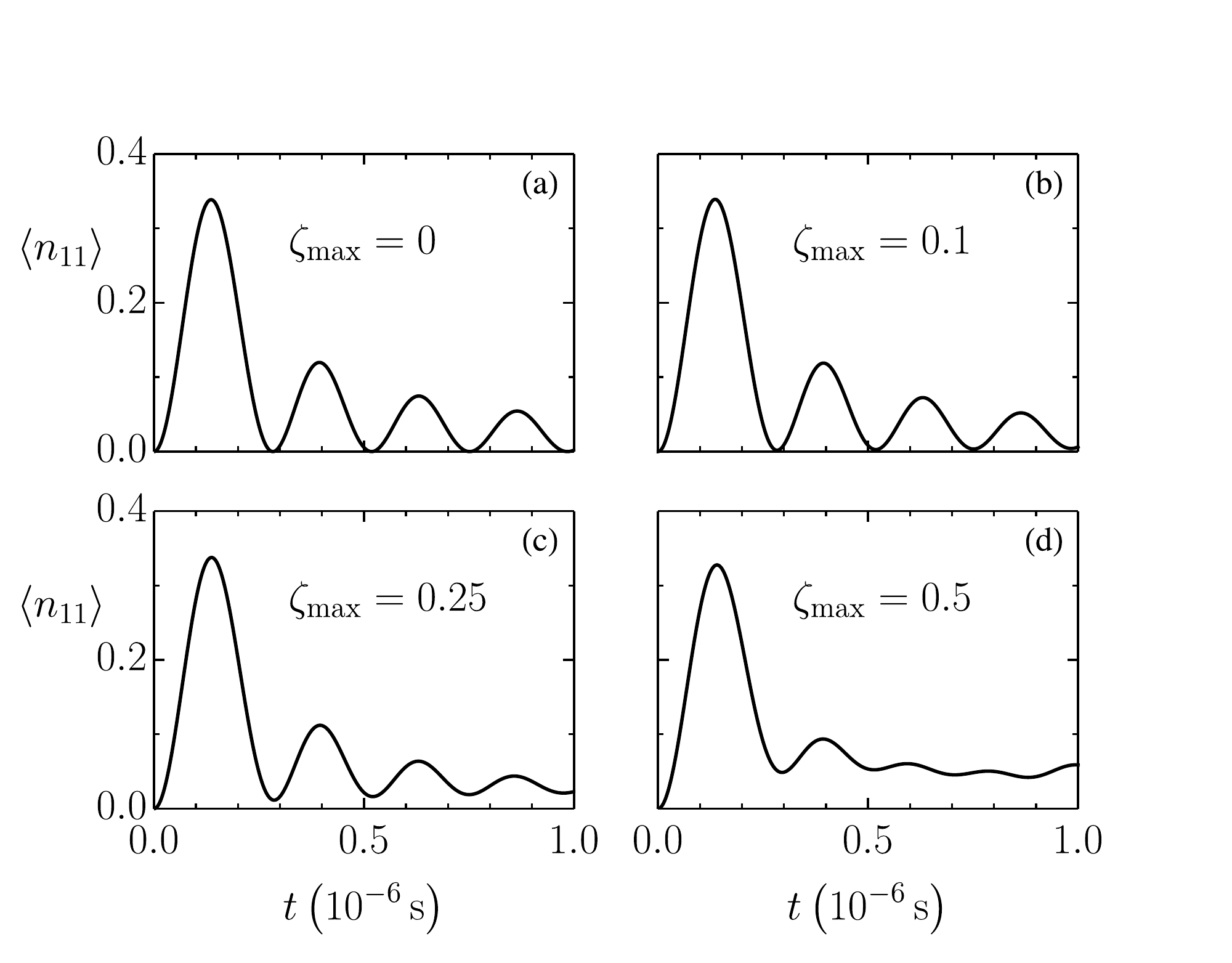}
		\caption{Evolution of the occupation of cavity 11 for disorder strengths (a) $\zetamax = 0$, (b) $\zetamax = 0.1$, (c) $\zetamax = 0.25$, and (d) $\zetamax = 0.5$.}
		\label{fig:cavnplot11}
	\end{center}
\end{figure}

For the single polariton calculations with the polariton initially in cavity 10, the simulations were run for $1\,\mu\second$ for the case of no disorder in the Rabi frequency and for the cases of uniform disorder with disorder strengths $\zetamax =$ 0.1, 0.25, and 0.5. For the disordered cases, we used 200 realizations of uniform disorder in the Rabi frequency and the polariton number in each cavity as a function of time was averaged over the 200 disorder realizations. The results obtained for the evolution of the polariton number in each cavity are shown in Figure~\ref{fig:mnplots}. Figure \ref{fig:mnplots} (a) shows the results in the absence of disorder, while Figures \ref{fig:mnplots} (b), \ref{fig:mnplots} (c), and \ref{fig:mnplots} (d) show the results for disorder strengths of $\zetamax =$ 0.1, 0.25, and 0.5 respectively. The results in Figure~\ref{fig:mnplots} (a) are qualitatively similar to the results of Petrosyan et al. for Bose-Hubbard dimers with no disorder~\cite{Petrosyan:2007rr}. Both without disorder and for the three disorder strengths, the wavefunction of the polariton disperses in both directions away from the initial cavity.

The evolutions of the number of polaritons in cavity 1 for the case of no disorder and for disorder strengths $\zetamax =$ 0.1, 0.25 and 0.5 are given in Figure~\ref{fig:cavnplot1}. The corresponding results for cavities 20, 10, and 11 are shown in Figure~\ref{fig:cavnplot20}, \ref{fig:cavnplot10}, and \ref{fig:cavnplot11} respectively. As shown in Figure~\ref{fig:cavnplot1}, the maximum occupation for cavity 1 occurs at roughly the same time for all disorder strengths, ranging from $0.84\,\mu\second$ for no disorder to $0.88\,\mu\second$ for $\zetamax = 0.5$. The maximum occupation obtained for cavity 1 decreases with increasing disorder strength, going from $0.272$ for no disorder to $0.186$ for $\zetamax = 0.5$. As shown in Figure~\ref{fig:cavnplot20}, the results are similar for cavity 20 where the maximum occupation occurs at times ranging from $0.92\,\mu\second$ for no disorder to $0.98\,\mu\second$ for $\zetamax = 0.5$. The maximum occupation for cavity 20 decreases from $0.260$ for no disorder to $0.168$ for $\zetamax = 0.5$. Figure \ref{fig:cavnplot10} shows that for cavity 10 in the absence of disorder, the occupation starts off at 1 and then decreases to 0 after $0.18\,\mu\second$. It then oscillates, periodically returning to 0, with the maxima decreasing with time. With disorder the behaviour is similar except that as the disorder increases, the oscillations are increasingly damped. The behaviour for cavity 11 shown in Figure~\ref{fig:cavnplot11} is similar to that of cavity 10, with the exception that the initial occupation is 0.

\begin{figure}
\begin{center}
\includegraphics[width=\linewidth]{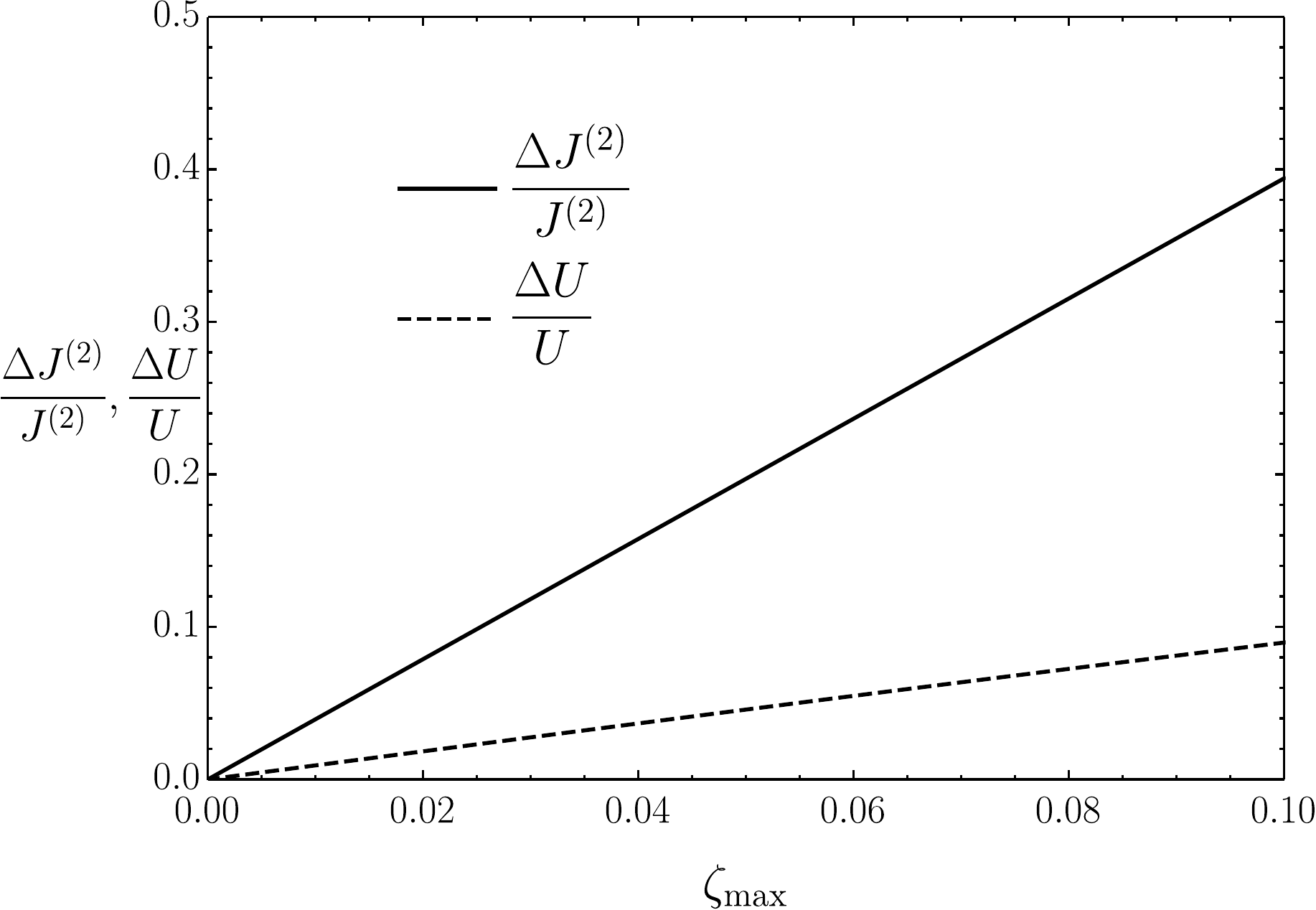}
\caption{Fractional disorder in $J^{(2)}$ and $U$ as a function of the disorder strength $\zetamax$.}
\label{fig:dJ2Uplot}
\end{center}
\end{figure}

For the 2 polariton calculations with both polaritons initially in cavity 10, the simulations were run for $10\,\mu\second$. This longer time was necessary for the wavefunction of the polaritons to disperse to the edges of the system in the absence of disorder. The simulations were run without disorder in the Rabi frequency and for 200 realizations of uniform disorder in the Rabi frequency with disorder strengths ranging from 0.005 to 0.3 in intervals of 0.005. In all cases, the projection of the wavefunction onto states, $\ket{2_i}$, with both polaritons in the same cavity was calculated. In the absence of disorder, this projection, given by $\sum_{i}\abs{\braket{2_i}{\Psi}}^2$, was found to be approximately 1 throughout the time evolution. This indicates that the two polaritons are strongly bound and we may speak of a polariton dimer. The presence of a dimer was also found for all disorder strengths considered. The dimer can be shown to hop with the effective second-order hopping rate $J^{(2)} = 2J^2/U$~\cite{Petrosyan:2007rr}. In the case considered in this paper we have $J^{(2)} \approx \quantity{6.2}{5}{\per\second} \ll J$. This smaller dimer hopping rate accounts for the need for a longer simulation time. 

With disorder, using the averaged values of  $J_{ij}$ and $U_{i}$ for each disorder strength, an effective ratio of $J^{(2)}$ to $U$ can be calculated as
\begin{equation}
\left.\frac{J^{(2)}}{U}\right\vert_{\zetamax} = 2\left.\left(\frac{\av{J_{ij}}}{\av{U_i}}\right)^2\right\vert_{\zetamax}. 
\end{equation}
Since the ratio $J/U$ is determined by the Rabi frequency, the ratio $J^{(2)}/U\vert_{\zetamax}$ in our simulations is approximately constant at 0.004 for any $\zetamax$. Figure \ref{fig:dJ2Uplot} shows the fractional disorder in the effective dimer hopping, $\Delta J^{(2)}/J^{(2)}$, and the fractional disorder in the on-site repulsion, $\Delta U/U$, as a function of the disorder strength. Here $\Delta J^{(2)}$ and $\Delta U$ represent the range of the values across the 200 disorder realizations for $J^{(2)}$ and $U$ respectively. Figure \ref{fig:dJ2Uplot} shows that the fractional disorder in the effective dimer hopping exceeds that in the on-site repulsion for any $\zetamax$.

\begin{figure*}
	\begin{center}
		\includegraphics[width=\linewidth]{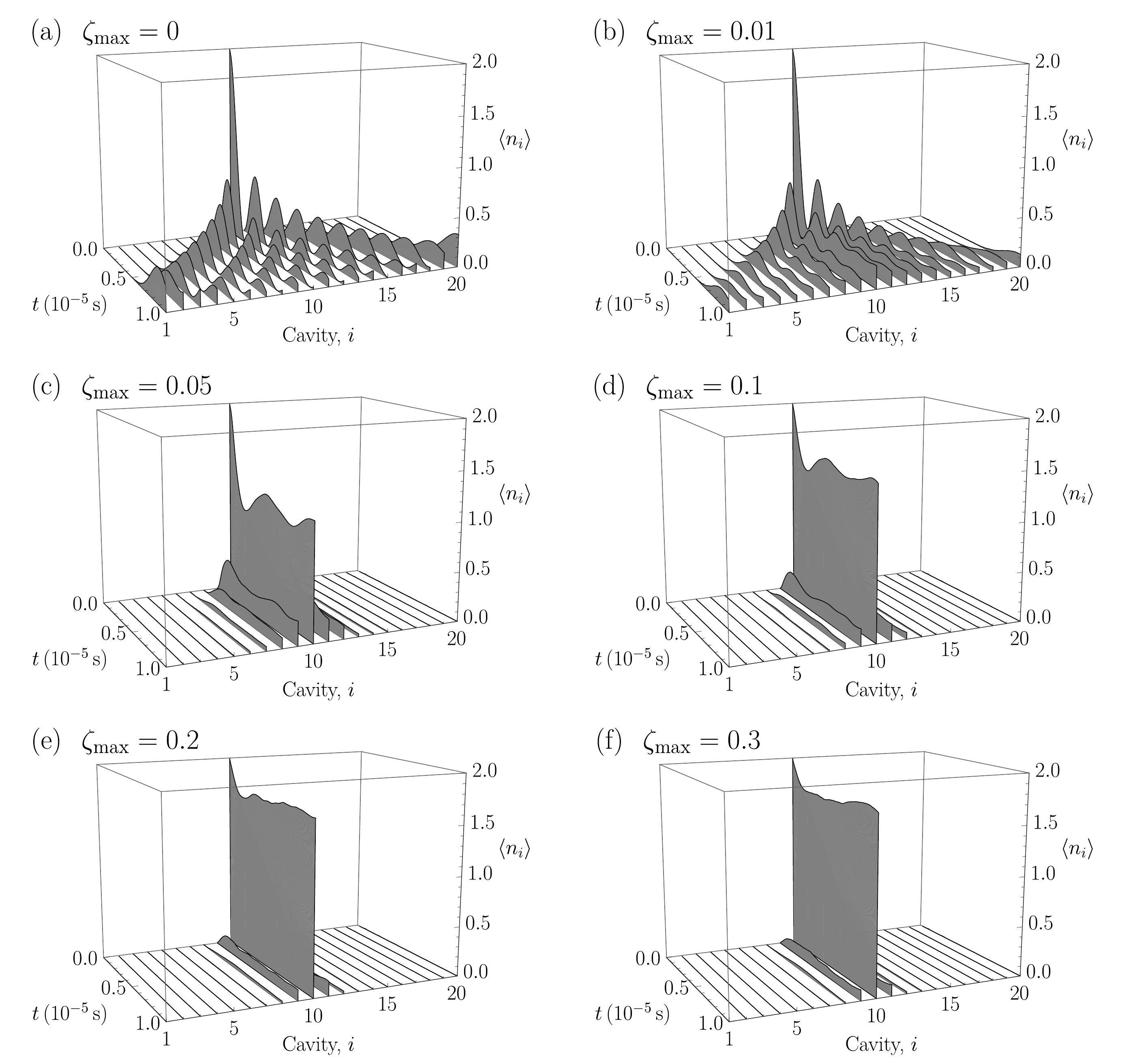}
		\caption{Dynamics of one dimer for disorder strengths (a) $\zetamax = 0$, (b) $\zetamax = 0.01$, (c) $\zetamax = 0.05$, (d) $\zetamax = 0.1$, (e) $\zetamax = 0.2$, and (f) $\zetamax = 0.3$.}
		\label{fig:nplots}
	\end{center}
\end{figure*}

Figure \ref{fig:nplots} shows the results for the averaged evolution of the number of polaritons in each cavity for the disorder strengths $\zetamax =$ 0, 0.01, 0.05, 0.1, 0.2, and 0.3. In the absence of disorder, the behaviour of the dimer is similar to the behaviour of the single polariton, with the wavefunction of the dimer dispersing in both directions away from the initial cavity, matching the results of Petrosyan et al.~\cite{Petrosyan:2007rr} However, unlike the case of the single polariton where the dispersal of the wavefunction is minimally affected by the addition of disorder, as the strength of the disorder increases, the dispersal of the dimer wavefunction is significantly reduced. Of the disorder strengths shown in Figure~\ref{fig:nplots}, we observe that there is localization of the dimer for $\zetamax \ge 0.05$. Analysis of the results for disorder strengths $\zetamax \ge 0.005$ indicated that there was localization of the dimer for disorder strengths of $\zetamax \ge 0.015$.

\begin{figure}
	\begin{center}
		\includegraphics[width=\linewidth]{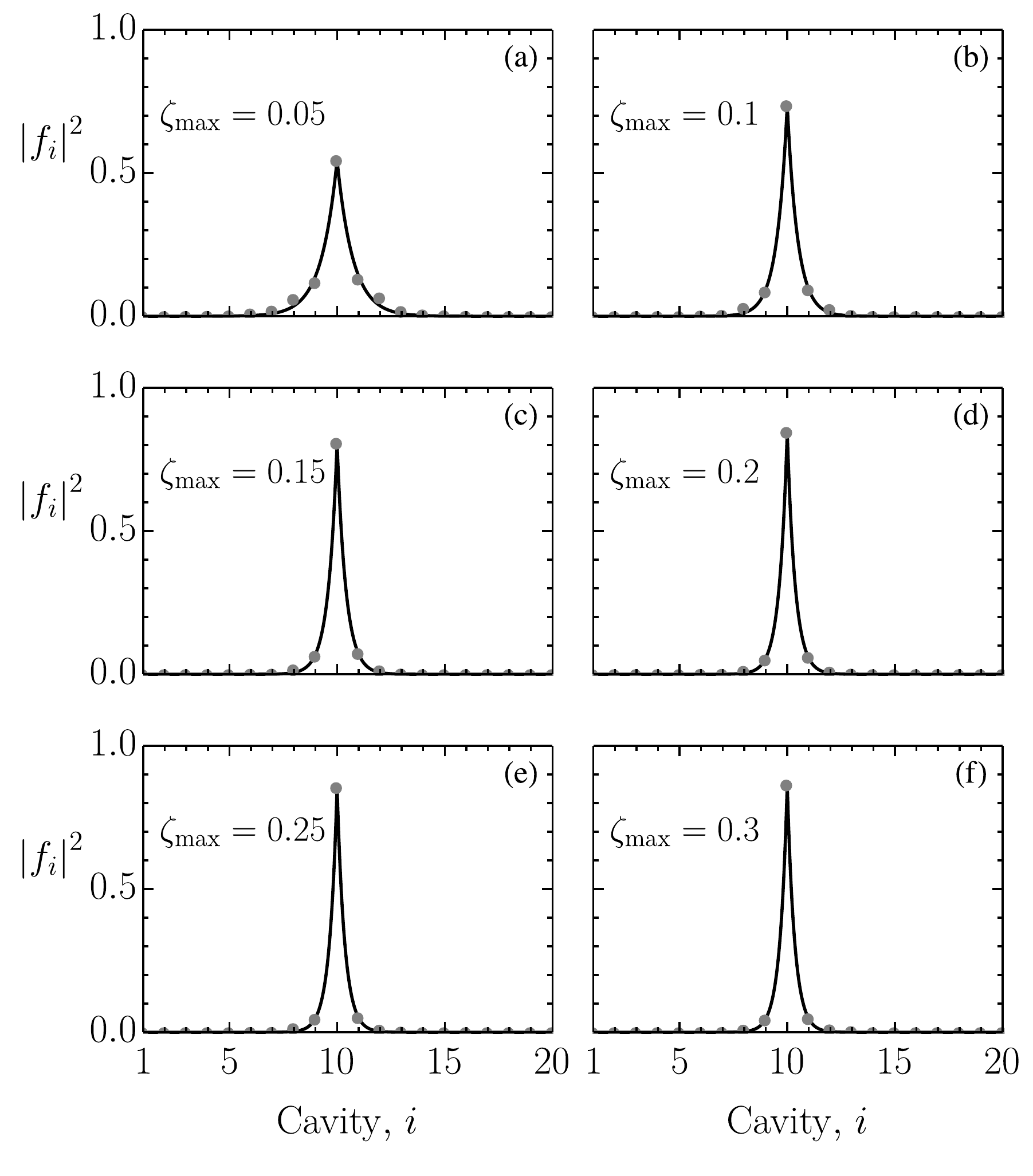}
		\caption{Exponential fit for $\abs{f_{i}}^2$ as a function of the cavity number, $i$.}
		\label{fig:fitplots}
	\end{center}
\end{figure}

\begin{figure}
	\begin{center}
		\includegraphics[width=\linewidth]{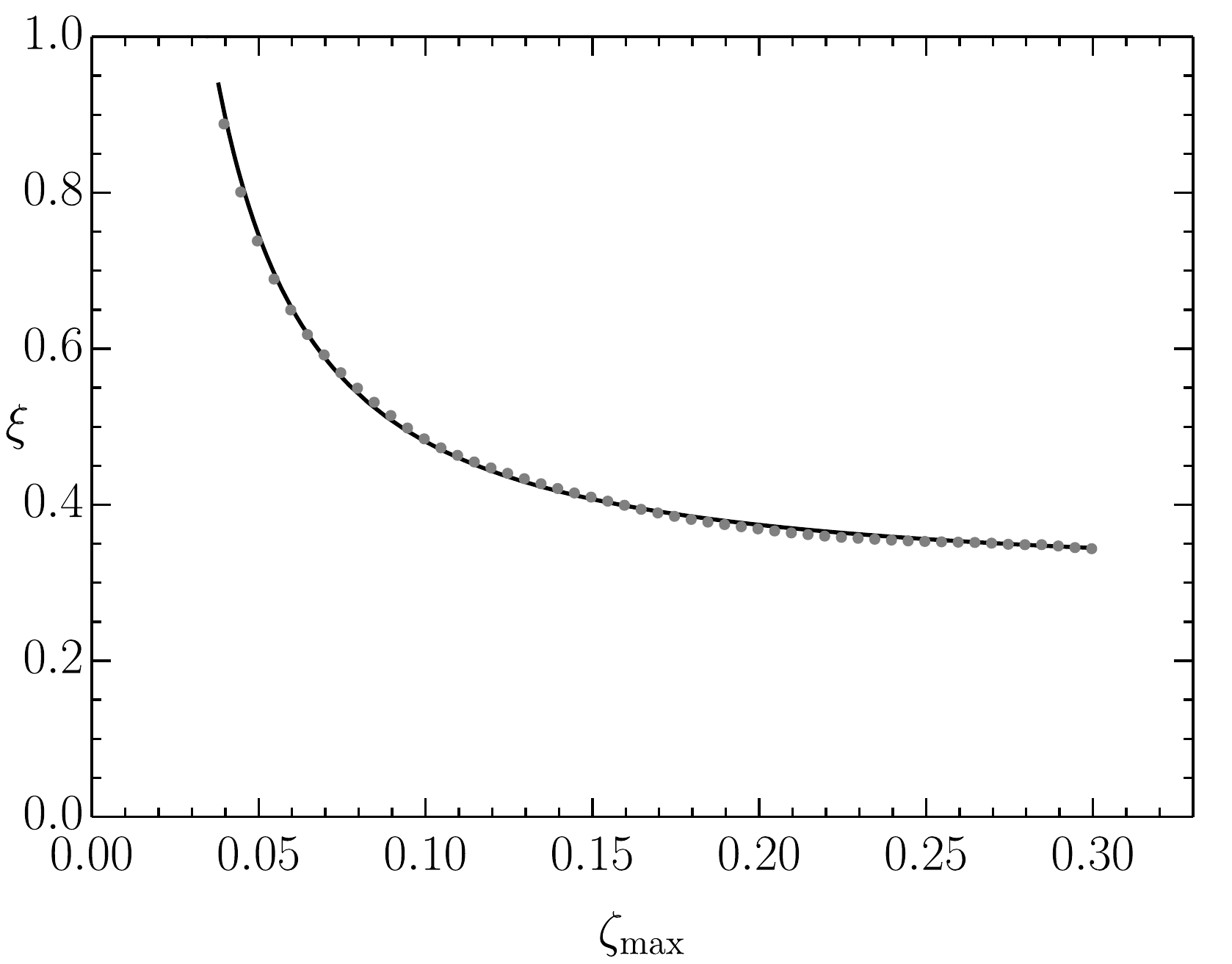}
		\caption{Localization length $\xi$ as a function of the disorder strength $\zetamax$ with a fit to the function $\xi(\zetamax) = a(\zetamax)^{-k} + b$.}
		\label{fig:llfitplot}
	\end{center}
\end{figure}

In order to quantify the localization we consider the evolution of the expansion coefficients of the dimer states. The wavefunction of the system can be expanded as 
\begin{equation}
\ket{\psi(t)} = \sum_{i}f_{i}(t)\ket{2_{i}} + \sum_{i< j}f_{ij}(t)\ket{1_{i}1_{j}}.
\end{equation}
As determined above from the projection, this state can be approximated as $\ket{\psi(t)} \approx \sum_{i}f_{i}(t)\ket{2_{i}}$. For each disorder strength between 0.02 and 0.3, we averaged the values of $\abs{f_{i}(t)}^2$ over the last $5\,\mu\second$ of the evolution and over the 200 disorder realizations to produce $\abs{f_{i}}^2$ for each cavity. The values of $\abs{f_{i}}^2$ were then fit with the function $\abs{f_{i}}^2=\abs{f_{10}}^2\ee^{-\abs{i-10}/\xi}$ to estimate the localization length $\xi$. Figure \ref{fig:fitplots} shows the fits obtained for $\zetamax = $ 0.05, 0.1, 0.15, 0.2, 0.25, and 0.3. The corresponding values of $\xi$ are 0.74, 0.49, 0.41, 0.37, 0.35, and 0.34. The localization length can therefore be seen to decrease as the disorder strength increases. Figure \ref{fig:llfitplot} shows the localization length plotted against the disorder strength, together with a fit with the function $\xi(\zetamax) = a(\zetamax)^{-k} + b$. The fit obtained was $\xi(\zetamax) = 0.0087(\zetamax)^{-1.31} + 0.302$. If we let the asymptotic value of $\xi(\zetamax)$ be $\xi_{\infty} = 0.302$, then $\xi(\zetamax) - \xi_{\infty}$ follows a decaying power law with exponent -1.31. The value of $\xi_{\infty}$ gives the localization length in the limit of infinite disorder. In all cases considered the localization length was less than one lattice constant.

\begin{figure}
\begin{center}
\includegraphics[width=\linewidth]{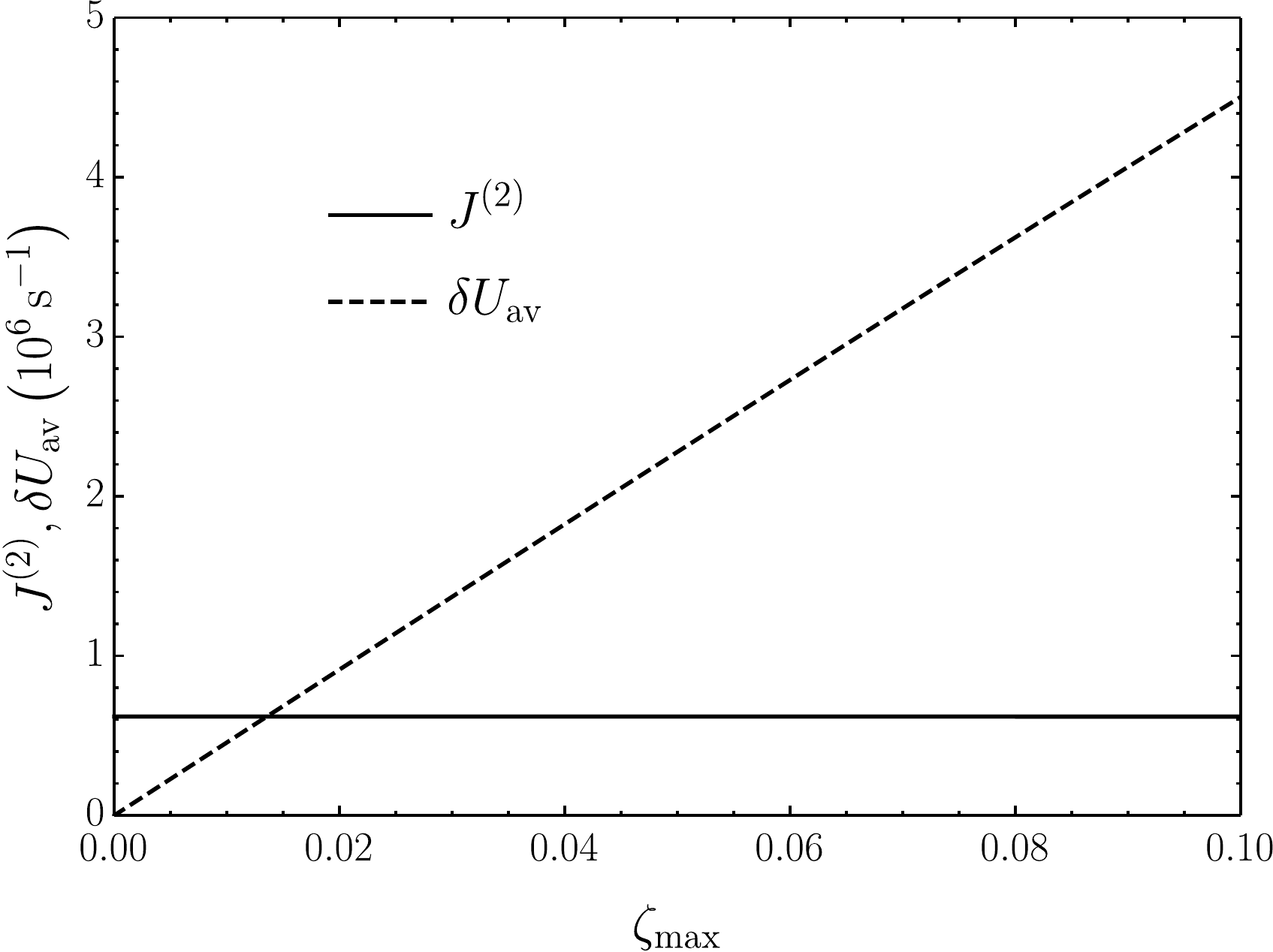}
\caption{$J^{(2)}$ and $\delta U_{\mathrm{av}}$ as a function of the disorder strength $\zetamax$.}
\label{fig:JDUplot}
\end{center}
\end{figure}

The difference between the results for the single polariton and the dimer can be seen as the consequence of the energy differences between different states. For a single polariton, the energy of all states, $\ket{1_i}$, with 1 polariton in a cavity is 0, regardless of the presence or strength of disorder. As a result, the hopping transition $\ket{1_i}\ket{0_{i+1}} \to \ket{0_i}\ket{1_{i+1}}$ does not change the energy of the system and there is no energy requirement for the hopping of the polariton to the nearest neighbour cavity. Therefore, it is energetically favourable for the polariton to move away from the initial cavity. On the other hand, for a dimer, the energy of a state, $\ket{2_i}$, is given by $U_i$, where $U_i$ is the on-site repulsion at cavity $i$. In the absence of disorder, $U_i = U$ for all cavities, $i$, and the transition $\ket{2_i}\ket{0_{i+1}} \to \ket{1_i}\ket{1_{i+1}} \to \ket{0_i}\ket{2_{i+1}}$ is resonant since the initial and final states have the same energy, $U$, and the transition conserves energy. Therefore, it is energetically favourable for the dimer to move away from the initial cavity in the absence of disorder. However, in the presence of disorder, $U_i$ varies from cavity to cavity and the energy difference between the states $\ket{2_i}\ket{0_{i+1}}$ and $\ket{0_i}\ket{2_{i+1}}$ is $U_{i+1} - U_i$, which is non-zero. As a result, the transition is off-resonant. Without dissipation or gain of energy, this off-resonant transition is suppressed due to a lack of energy conservation. On average, the energy difference between adjacent cavities, $\abs{U_{i+1} - U_i}$, increases as the disorder strength increases, thus the dispersal of the dimer wavefunction is suppressed more for larger disorder strengths. Figure \ref{fig:JDUplot} shows the effective dimer hopping rate, $J^{(2)}$, and the average of the magnitude of the disorder in $U$ between adjacent sites, $\delta U_{\mathrm{av}}$, as a function of the disorder strength $\zetamax$. The quantity $\delta U_{\mathrm{av}}$ was calculated from the disorder realizations using $\delta U_{\mathrm{av}} = \av{\abs{U_{i+1} - U_{i}}}$. The point at which $J^{(2)} = \delta U_{\mathrm{av}}$ is approximately $\zetamax = 0.014$. This agrees with the value obtained from the numerical simulations for the critical value of the disorder strength at which localization begins.

\begin{figure*}
	\begin{center}
		\includegraphics[width=\linewidth]{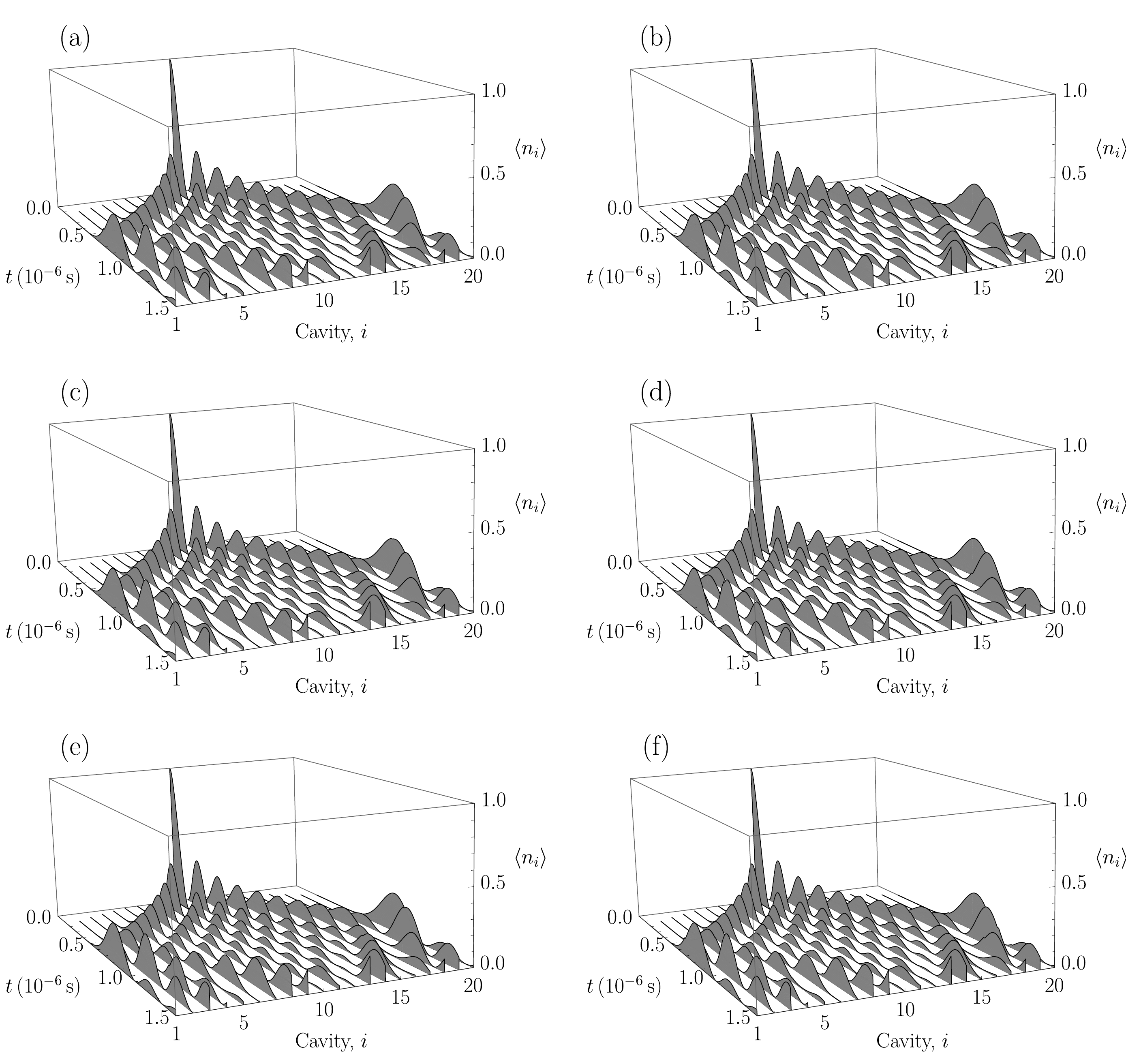}
		\caption{Dynamics of one polariton with uniform disorder of strength $\zetamax = 0.1$ being switched on at (a) $0.2\,\mu\second$, (b) $0.3\,\mu\second$, (c) $0.4\,\mu\second$, (d) $0.5\,\mu\second$, (e) $0.6\,\mu\second$, and (f) $0.7\,\mu\second$, and then switched off after $0.5\,\mu\second$.}
		\label{fig:switchmnplots}
	\end{center}
\end{figure*}

We now examine the effect of instantaneously turning on and then later turning off the disorder in the Rabi frequency. For the case of a single polariton, we ran 6 different simulations for a time period of $1.5\,\mu\second$, with disorder being switched on in each simulation at $0.2\,\mu\second$, $0.3\,\mu\second$, $0.4\,\mu\second$, $0.5\,\mu\second$, $0.6\,\mu\second$, and $0.7\,\mu\second$ respectively. In each simulation, the disorder was switched back off after $0.5\,\mu\second$ had elapsed. The simulations were run for 200 realizations of uniform disorder in the Rabi frequency with disorder strength $\zetamax = 0.1$, and the number of polaritons in each cavity was averaged over the 200 realizations. The results for the 6 disorder start times are shown in Figure~\ref{fig:switchmnplots}. The results show that turning the disorder on and off has no effect on the dispersal of the polariton wavefunction from the initial site. This is as expected since we have established that disorder has little effect on a single polariton.

\begin{figure*}
	\begin{center}
		\includegraphics[width=\linewidth]{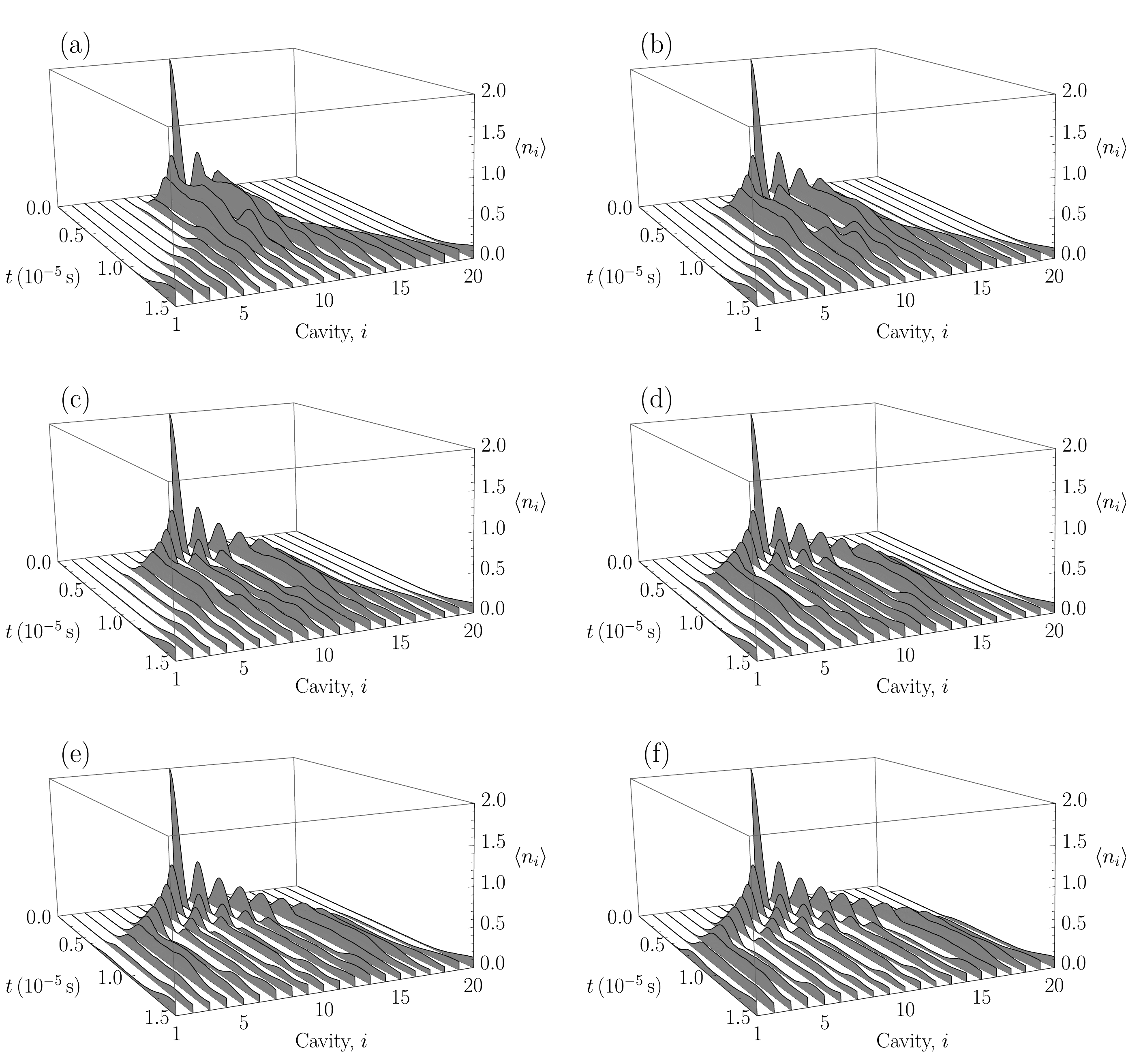}
		\caption{Dynamics of one dimer with uniform disorder of strength $\zetamax = 0.1$ being switched on at (a) $2\,\mu\second$, (b) $3\,\mu\second$, (c) $4\,\mu\second$, (d) $5\,\mu\second$, (e) $6\,\mu\second$, and (f) $7\,\mu\second$, and then switched off after $5\,\mu\second$.}
		\label{fig:switchnplots}
	\end{center}
\end{figure*}

The same simulations were run for the case of a single dimer using a time period of $15\,\mu\second$, with the disorder switching times increased by a factor of 10. Figure \ref{fig:switchnplots} shows the results for the 6 disorder start times. The results show that while the disorder is switched off, the dimer wavefunction disperses from the initial cavity in both directions. When the disorder is switched on, the dispersal stops and the dimer wavefunction becomes trapped. When the disorder is later switched off, the dimer wavefunction begins dispersing again. These simulations demonstrate the ability to control the propagation of the dimer away from the initial cavity by switching on and off the disorder in the Rabi frequency in a controlled manner.

\section{Conclusion}
\label{sec:Conclusion}

In this paper, we examined the effect of disorder in the Rabi frequency of the driving laser on the dynamics of polaritons and dimers in a coupled array of optical cavities. We compared the behaviour of a polariton dimer to the behaviour of a single polariton. We found that the addition of disorder in the Rabi frequency had little effect on the dynamics of a single polariton in an array of 20 cavities with the polariton wavefunction dispersing away from the initial cavity both with and without disorder. For all the disorder strengths considered, the maximum occupation in cavities 1 and 20 occurred at roughly the same time, with the maximum occupation decreasing as the disorder strength increased. For cavities 10 and 11, the polariton occupation oscillated, with the oscillations being increasingly damped as the disorder strength increased.

In contrast to the single polariton, the addition of disorder caused localization of the polariton dimer, with the localization length decreasing as the disorder strength increased. The critical value of the disorder strength for localization to occur was found to be approximately 0.015. After subtracting an asymptotic value of 0.302, the localization length was found to follow a decaying power law as a function of the disorder strength with an exponent of -1.31.

We also investigated the effect of switching on and off the disorder in the Rabi frequency at different times during the evolution. In the case of a single polariton, switching on and off the disorder had no effect on the dispersal of the polariton wavefunction. In the case of the polariton dimer, however, the dimer wavefunction initially disperses from the initial cavity while the disorder is off. Switching on the disorder halts the dispersal of the dimer wavefunction, and later switching off the disorder allows the wavefunction to continue dispersing again. Therefore, the localization of the dimer can be controlled temporally by switching on the disorder at any given time during the evolution process.

\section*{Author contribution statement}

Abuenameh Aiyejina performed the simulations. Both authors wrote the manuscript together.

\bibliographystyle{epj}
\bibliography{paper}   

\end{document}